\documentclass[aps,prl,twocolumn,showpacs]{revtex4}
\usepackage{amssymb}
\usepackage{amsmath}
\usepackage{graphicx}

\begin{document}
\title{Realization of a cascaded quantum system: heralded absorption of a single photon qubit by a single-electron charged quantum dot}
\author{Aymeric Delteil}
\author{Zhe Sun}
\author{Stefan F\"alt}
\author{Atac Imamo\u{g}lu}

\affiliation{Institute of Quantum Electronics, ETH Zurich, CH-8093
Zurich, Switzerland.}

\date{\today }

\begin{abstract}
Photonic losses pose a major limitation for implementation of quantum state transfer between nodes of a quantum network. A measurement that heralds successful transfer without revealing any information about the qubit may alleviate this limitation. Here, we demonstrate heralded absorption of a single photonic qubit generated by a single neutral quantum dot, by a single-electron charged quantum dot that is located 5 meters away. The transfer of quantum information to the spin degree of freedom takes place upon emission of a photon: for a properly chosen or prepared quantum dot, detection of this photon yields no information about the qubit. We show that this process can be combined with local operations optically performed on the destination node, by measuring classical correlations between the absorbed photon color and the final state of the electron spin. Our work suggests alternative avenues for realization of quantum information protocols based on cascaded quantum systems.

\end{abstract}

 \maketitle

In the context of quantum communication and distributed quantum computing, the ability to faithfully transfer a quantum state from one node to another is of key importance~\cite{DiVincenzo00}. In a direct transfer scheme, quantum information would in principle be transmitted between nodes using consecutive emission, propagation and absorption of single optical photons~\cite{Cirac97,Rezus12,Meyer15,Piro10}.
However realization of such a full procedure is most challenging in any physical system, due to inefficient photon collection from single-photon sources as well as propagation losses. Previous realization of such transfer using atoms was achieved with low efficiency and without an experimental verification of successful transfer~\cite{Ritter12}. On the other hand, a heralding signal, provided for instance by detection of a subsequently emitted photon at a different wavelength, can make the transfer process robust to losses in protocols where multiple attempts are possible. Previous demonstrations of heralded photon-to-matter state transfer with atoms or NV centers were implemented using classical light pulses from weak lasers~\cite{Kurz14,Kalb15,Yang16} instead of true single photon pulses generated by a single quantum system. A heralded transfer of quantum information from a single flying qubit to a single stationary qubit has so far remained elusive.

\begin{figure}[h]
  \centering
  \includegraphics[width=3.5in]{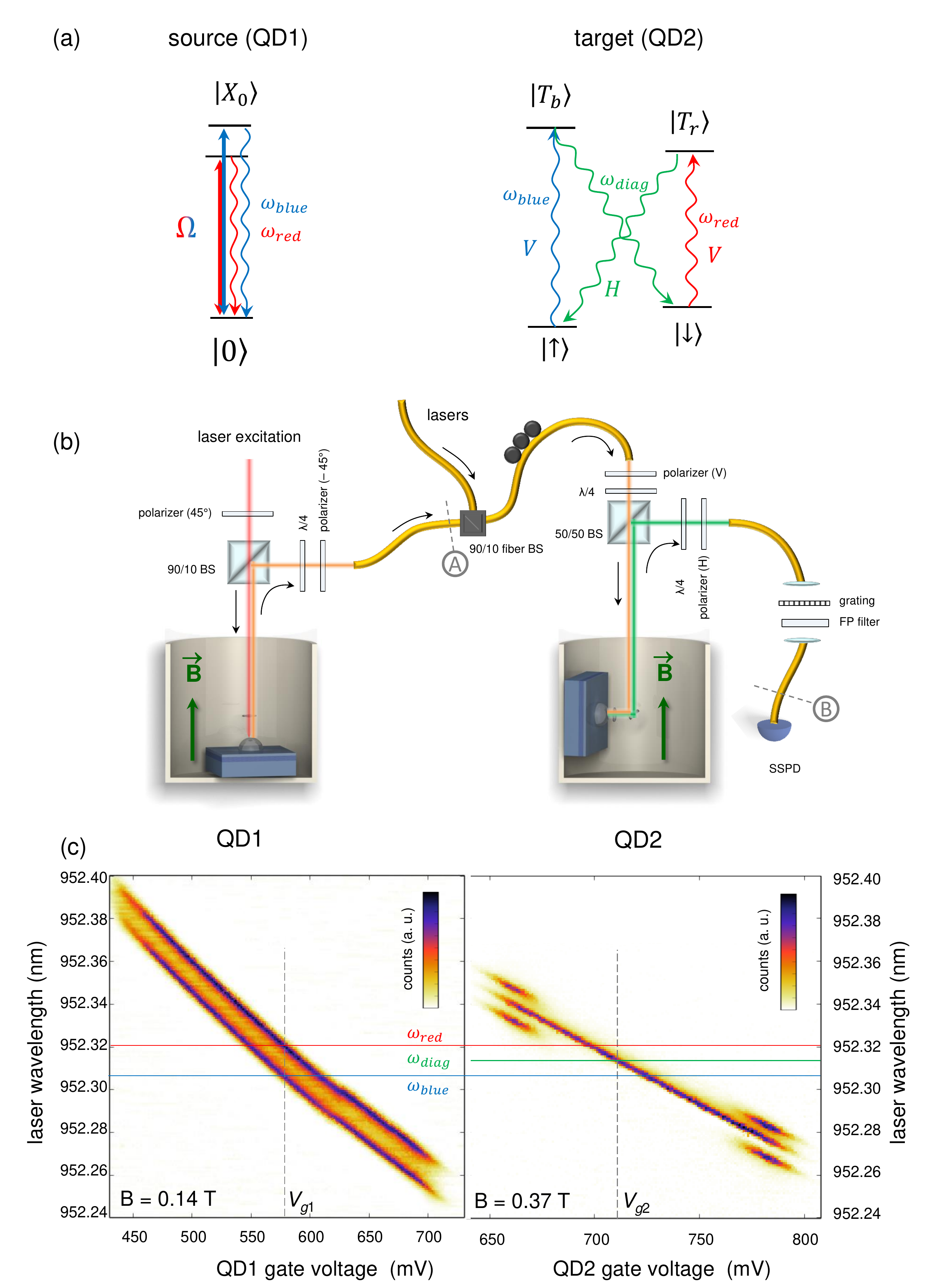}\\
  \caption{(a) Energy levels of the source (left) and target (right) QDs. (b) Experimental set-up. The neutral exciton transitions of a QD (QD1, left cryostat) are resonantly excited by a two-color laser. The spontaneously emitted photons are collected into a fiber and conveyed to the input arm of a second cryostat, which hosts a second QD (QD2) in the single electron charged state, with an external magnetic field applied in the Voigt geometry. The scattered photons are once more collected into a fiber, filtered by a grating and, when applicable, a Fabry-P\'erot (FP) filter and detected with a superconducting single photon detector (SSPD). (b) RF plateau scans of QD1 and QD2 at the magnetic fields used in the experiments. The vertical axis (laser wavelength) is the same for both plots. The gate voltages used in the experiments is indicated with the vertical dashed line, and the horizontal lines indicate the relevant transition wavelengths.}\label{1}
\end{figure}

In this Letter, we demonstrate heralded absorption of a single photonic qubit by a single-electron charged quantum dot (QD). Our work constitutes an experimental realization of a cascaded quantum system~\cite{Carmichael93,Gardiner93}, where fluorescence from a source quantum emitter (a neutral QD) is used to drive a second target quantum emitter (a single electron charged QD); the detection of a photon from the latter heralds the success of single photon absorption. Furthermore, we show that the color of the photon generated by the source QD is classically correlated with the spin state of the electron in the target QD upon completion of the transfer process. The scheme we realize requires that the source and the target QD have energy level diagrams depicted in Figure~\ref{1}a: the neutral exciton fine structure allows for the possibility to generate a photonic color qubit in an arbitrary superposition of two center frequencies $\omega_{blue}$ and $\omega_{red}$ \cite{Gao13}. Degeneracy of the diagonal transitions of the target QD on the other hand, is essential for heralding the absorption of the incoming photonic qubit in state $\alpha |\omega_{red} \rangle + \beta |\omega_{blue} \rangle$, without revealing any information about the qubit state. Provided that the electron spin is initially prepared in a superposition state $\left( |\uparrow \rangle +  |\downarrow \rangle \right)/\sqrt{2}$, the absorption of the incoming photon and the subsequent emission of a diagonal photon at frequency $\omega_{diag}$ projects the electron spin onto the state $\alpha |\uparrow \rangle + \beta |\downarrow \rangle$. By choosing a single-electron charged source QD, this scheme can be modified to obtain spin-to-spin quantum state transfer or heralded entanglement generation (see Supplemental Material~\cite{SOM}).

Self-assembled QDs are one of the most favorable systems for building efficient quantum links between remote nodes using optical photons~\cite{Gao13,Delteil15}, due to their integrability into microcavities that allow efficient photon collection into fibers~\cite{Gazzano13, Munsch13} and near-unity quantum efficiency of emission into a zero-phonon-line. In this work, we use two identical samples of single self-assembled InGaAs/GaAs QDs embedded in a lossy planar cavity consisting of a bottom thick (28~periods) distributed Bragg mirror (DBR) and a top thin (2~periods) DBR. A p-i-n diode structure is grown around the QD layer for charge state control and Stark tuning of the transitions. A ZnO solid immersion lens (SIL) is mounted on top of the samples to increase the collection efficiency. The samples are held in two helium bath cryostats with confocal microscopes allowing to excite QDs with lasers and collect scattered photons through the same objective.

The experimental set-up is depicted Fig.~\ref{1}b: single photon pulses are generated through resonant optical excitation of a neutral QD (labeled QD1). By making use of the exciton fine structure, single photons can be generated in any superposition of two center frequencies $\omega_{blue}$ and $\omega_{red}$ \cite{Gao13}. They are conveyed via a fiber to another QD located 5~meters away (QD2), which is charged with a single electron. An external magnetic field is applied to QD2 in Voigt geometry, making all four transitions from the two trion states to the two spin ground states allowed, with equal oscillator strength. Moreover, QD2 has been selected to have its two diagonal transitions almost degenerate, which is ensured by opposite in-plane $g$-factors for the electron and the hole. Although such QDs can be naturally found in self-assembled QD samples (see Supplemental Material~\cite{SOM}), the diagonal transitions can also be tuned to resonance using external electric field~\cite{Pretchel15}, strain~\cite{Tholen16}, ac-Stark shift~\cite{Bose12} or magnetic field orientation using a vector-magnet. 

 Fig.~\ref{1}c shows resonance fluorescence (RF) plateaus of the two QDs. While QD1 RF exhibits standard fine-structure-split lines of the neutral exciton, QD2 RF exhibits unique signature of degenerate diagonal transitions: instead of four RF lines visible only near the edges of the plateau due to spin pumping~\cite{Xu07}, only three lines are visible. Moreover, the intensity of the intermediate RF line does not vanish in the middle of the gate voltage range since a laser at $\omega_{diag}$ ensures that there is no spin pumping.  The external electric and magnetic fields are adjusted such that the two transitions of QD1 match the blue and red vertical transitions of QD2, as indicated in Fig.~\ref{1}c. For QD2, we have carried out Hong-Ou-Mandel interferometry to verify the indistinguishability of photons emitted at $\omega_{diag}$ (see Supplemental Material~\cite{SOM}).


The first part of our demonstration consists in exciting QD1 with a two-color laser in cw, generated from a single-mode laser using an electro-optic modulator (EOM) driven with a microwave signal at $(\omega_{blue} - \omega_{red})/2$. The resonant fluorescence photons are brought to the second cryostat, and their polarization is made vertical ($V$). The photons that are scattered a second time by QD2 pass through a horizontally ($H$) oriented polarizer that suppresses reflected QD1 light, such that  we collect only diagonal photons at frequency $\omega_{diag}$. The collection path also includes a grating of bandwidth $\sim 80$~GHz and a flip-mounted Fabry-P\'erot filter (that we do not use in this part of the experiment). Finally, we placed a superconducting single-photon detector (SSPD) at the output port of the collection fiber. Its dark count rate is lower than 1~s$^{-1}$, and so is the background light detection rate, leading to a false events recording rate lower than 2~s$^{-1}$, which is essential for the experiments presented here.

\begin{figure}[h]
  \centering
  \includegraphics[width=3.5in]{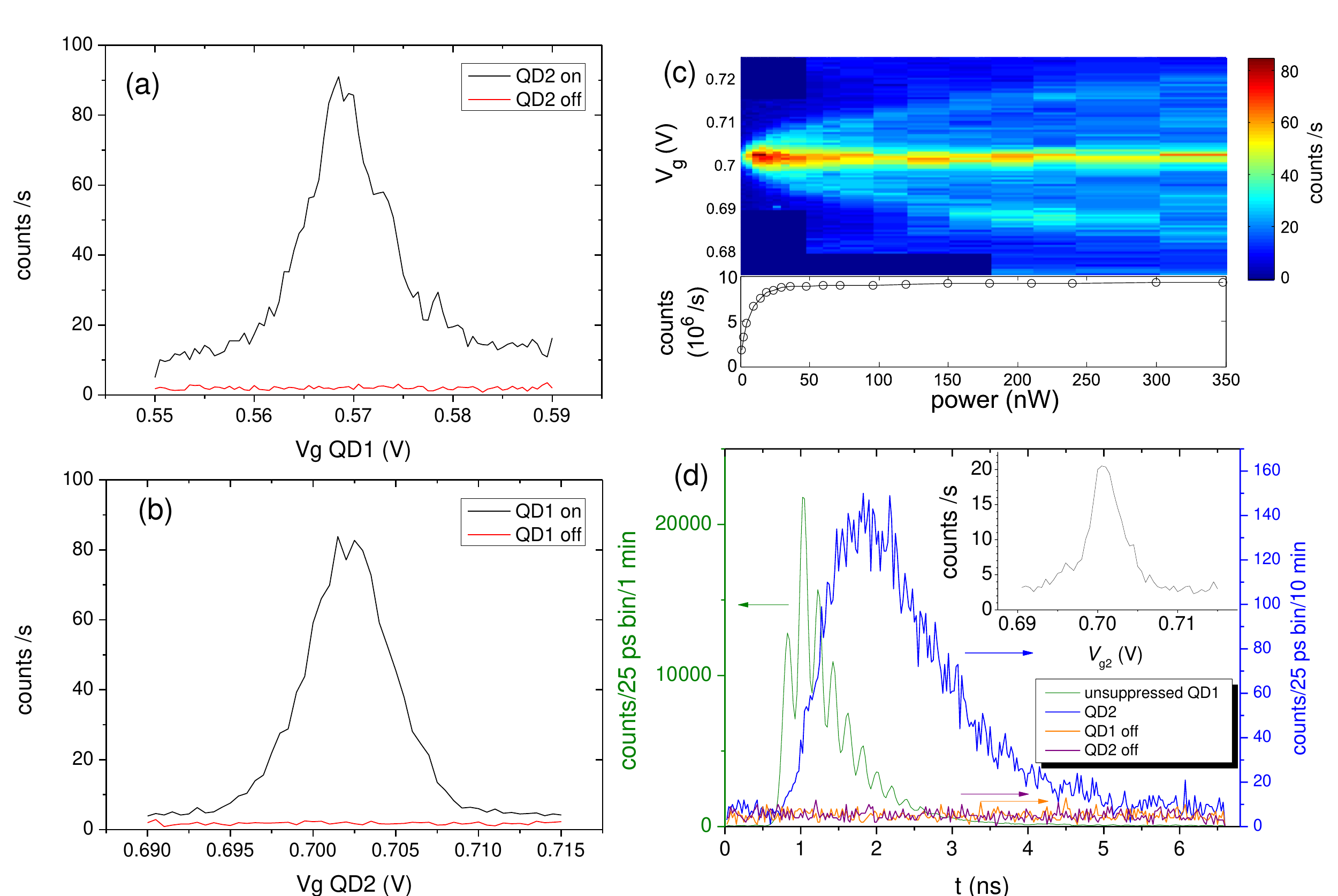}\\
  \caption{(a) Photon count rate as a function of QD1 gate voltage, with QD2 fixed on resonance (black curve) or off resonance (red curve). The integration time is 10~s per data point. (b) Photon count rate as a function QD2 gate voltage, with QD1 fixed on resonance (black curve) or off resonance (red curve). (c) Upper panel: count rate as a function of QD1 gate voltage and laser power, revealing the Mollow sidebands at high power. Lower panel: corresponding count rate of QD1 measured directly at the output fiber of the left cryostat (point~$A$ on Fig.~\ref{1}a), as a function of the laser power. (d) Time-resolved count rate, with a repetition rate of 7~ns: green curve, without cross-polarization suppression, revealing the single-photon pulse envelope from QD1. Blue curve: with cross-polarization suppression, showing the time-resolved heralded absorption rate. Orange (purple) curve: same as the blue curve, but when QD1 (QD2) is tuned off resonance. Inset: average count rate as a function of QD2 detuning.}\label{2}
\end{figure}

When we fix the laser wavelength and the gate voltage of QD2 at resonance, the count rate shows a Lorentzian line shape as a function of QD1 gate voltage, with a maximum of about 90~counts per second (Fig~\ref{2}a). Fixing QD1 and varying QD2 leads to a similar result (Fig~\ref{2}b). When either dot is tuned off resonance, the count rate remains as low as 2~s$^{-1}$, close to the background value, irrespectively of the gate voltage of the other dot (red curves of Fig~\ref{2}a and Fig~\ref{2}b), demonstrating unequivocally that we have implemented heralded single photon absorption. For these measurements, the laser power is set close to the saturation power of QD1, corresponding to the detection of about $5.5\cdot10^6$~counts/s at the output fiber of the first cryostat (point $A$ on Fig.~\ref{1}a). Although QD1 can be driven at higher power, leading to a detection rate up to almost 10~million counts/s, it does not result in an increase of the heralded absorption rate. Indeed when increasing the power above saturation, the photons emitted by QD1 start having a spectral mismatch with QD2 transitions, attributed to the appearance of Mollow sidebands~\cite{Vamivakas09,Flagg09}. These sidebands can be directly observed by means of the heralded scattering rate as a function of the gate voltage of QD2 as can be seen on Fig.~\ref{2}c.

The quantum efficiency of the heralded absorption process can be estimated knowing the total transmission of our set-up ($\sim$ 0.3~\%, measured from $A$ to $B$ on Fig.~\ref{1}a) and taking into account the collection efficiency of the scattered photons, including the collection at the first lens (20~\%, deduced from the maximum count rate and the measured excited state lifetime) and the polarizer (blocking 50~\% of the scattered photons). We deduce that about 8~\% of the incident photons are absorbed by QD2. This experiment is done with random spin population of QD2, leading to an additional reduction of 50~\% in this probability -- the quantum efficiency is therefore estimated to be about 16~\%. This number is consistently close to the overlap of QD2 emission with the free-space mode collected by the microscope objective ($\sim 20$~\%) which gives the maximum absorption probability of a single incident photon by a single quantum emitter~\cite{Pinotsi08}.

This experiment can also be performed in pulsed regime, allowing to observe the dynamics of the process within a few tens of picoseconds (Fig.~\ref{2}d). In this case, QD1 is driven by a two-color laser pulse of 400~ps, generating two-color single photon pulses with a repetition rate of 152~MHz. They can be directly detected by removing the cross-polarization suppression (green curve of Fig.~2d). The time-resolved detection rate exhibits a beat note at $\omega_{blue}-\omega_{red} = 4.9$~GHz, whose visibility is limited by the jitter of our detector. The $g^{(2)}(0)$ of the single-photon pulses is measured to be $0.15 \pm 0.02$, limited by double excitations during the 400~ps laser pulse. When the crossed polarizer is set, the photons at frequency $\omega_{diag}$ scattered by QD2 can then be observed (blue curve in Fig.~\ref{2}d). We also verify once more that the counts vanish when either dot is switched off-resonance (orange and purple curves on Fig.~\ref{2}d).

The time-resolved detection of these doubly scattered photons well above the background opens the possibility to implement quantum information protocols based on such cascaded quantum systems, by combining heralded absorption of photonic qubits with local operations such as state initialization, manipulation and readout. In the following, we demonstrate an experimental implementation of a photon-to-spin state transfer protocol, and we show classical correlations between the photonic qubit state and the final spin state.

\begin{figure}[h]
  \centering
  \includegraphics[width=3.5in]{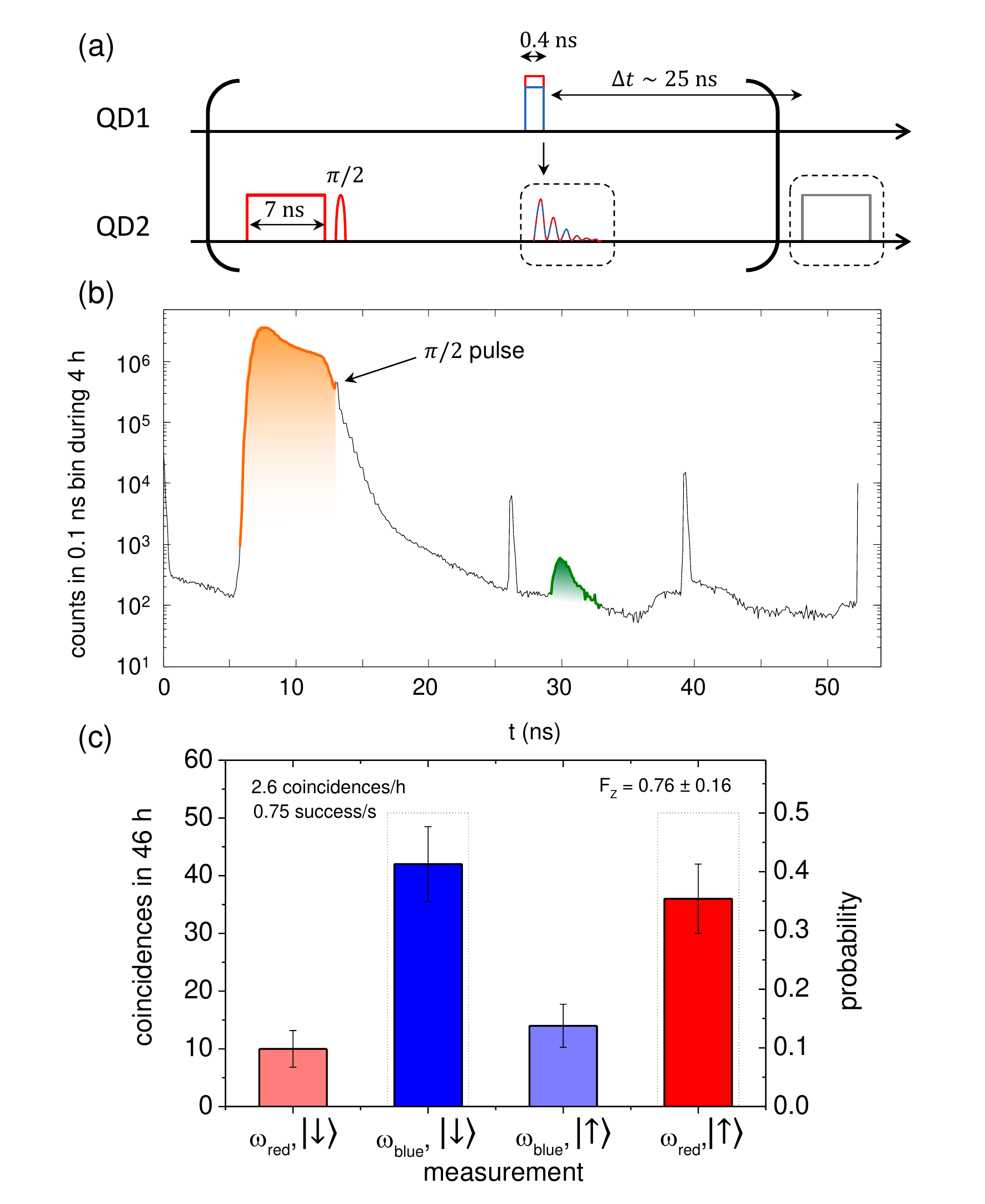}\\
  \caption{(a) Pulse sequence used to measure classical correlations. Single photon pulses are generated from QD1 using a 0.4~ns pulse, resonant with either of the exciton resonances. QD2 is prepared in state $(|\uparrow \rangle + |\downarrow \rangle)/ \sqrt{2}$ using spin pumping followed by a $\pi/2$ rotation. These pulses are combined with the single photon pulse generated using QD1.  (b) Photon count histogram recorded during a 4~hour long measurement. The events involved in the two-fold coincidence calculation are shaded in orange (spin measurement) and in green (heralded absorption of a single photon from QD1). The peaks at 26, 39 and 52~ns are residual ps-pulses from the rotation laser that are suppressed by $\sim 10^2$ using an EOM. (c) Two-fold coincidence rate corresponding to a photon detected during the spin pumping pulse conditioned on a photon heralding a single photon absorption in the previous period. The dashed bars correspond to the case of perfect classical correlations.}\label{3}
\end{figure}

The set-up for this experiment is the same as depicted in Fig.~\ref{1}a~\cite{SOM}.  The pulse sequence is shown Fig.~\ref{3}a. The spin state of QD2 is prepared in a superposition of equal weights by spin pumping in $|\uparrow \rangle$ or $|\downarrow \rangle$ followed by a rotation of $\pi/2$ performed using a 15-picosecond laser pulse detuned by 1~THz from the trion resonances. This rotation requires the insertion of a quarter-wave plate in the input arm such that the excitation port is elliptically polarized~\cite{Press08}. As a consequence, the crossed-polarized detection path is also elliptically polarized and therefore does not reject all the $V$-polarized blue and red photons scattered by QD2. They are then filtered out by adding a Fabry-P\'erot filter in the collection path, having one of the transmission maxima centered at $\omega_{diag}$. A single photon pulse, generated by the excitation of QD1 exciton, is coupled to QD2 after a delay time of $\sim 15$~ns, chosen such that the trion population has sufficiently decayed to observe heralded absorption above the spontaneous emission background. After an additional time of 25~ns (corresponding to the detector dead time), the pulse sequence is repeated, and the subsequent spin pumping pulse serves to measure the spin state in the computational basis states. As we do not perform dynamical decoupling, the spin superposition precesses with a random frequency during the protocol. Since in this work we demonstrate only classical correlations, this random phase plays no role in the obtained results. A detailed scheme of the experimental set-up including generation and synchronization of the optical pulses is available in the Supplemental Material~\cite{SOM}.

Fig~\ref{3}b shows a time trace recorded during four hours. One of the major technical difficulties is to keep a low background -- at least in the vicinity of the time window where we detect heralded absorption -- when combining strong resonant laser pulses of $P \sim P_{sat}$ with the single photon pulses. As usual EOMs have finite on/off ratios (typically $10^2$ to $10^3$) leading to a constant RF background of a few thousands of counts per second (for a dot showing a few $10^6$ counts per second around saturation), we cascade three EOMs to reach on/off ratio of $10^6$. This in turn suppresses the scattered photon background and allows for observation of heralded absorption (see Supplemental Material~\cite{SOM} for more details about contributions to the background).

The four combinations of input photon color (red and blue) and spin state measurement ($|\uparrow \rangle$ and $|\downarrow \rangle$) are alternated, and classical correlations between the photon color and the spin state are deduced from the two-fold coincidences of a first photon detected during the heralded absorption window (green shading on Fig.~\ref{3}b) together with a photon detected during the following spin pumping pulse (orange shading on Fig.~\ref{3}b). Using a repetition rate of 20~MHz, we obtain a success rate (rate of heralded absorption events) of 0.75~s$^{-1}$, corresponding to the count rate in the green shaded time window in Fig.~\ref{3}b. It is smaller than the rate shown in Fig.~\ref{2}d due to the much lower repetition rate, as well as the use of elliptic polarization in both input and output arms and the FP filter transmission. This leads to a two-fold coincidence rate of 2.6~per hour. The result from data recorded during 46~h is shown in Fig.~\ref{3}c. For a red (blue) input photon, the spin is measured in $|\uparrow \rangle$ ($|\downarrow \rangle$) $\sim$~3.5 times more often than in $|\downarrow \rangle$ ($|\uparrow \rangle$), consistent with what is expected (cf. QD2 level scheme on Fig.~\ref{1}a).  It leads to a computation basis fidelity of $0.76 \pm 0.16$, mainly limited by the background photons during the heralded absorption process.

Our experiments implement a cascaded quantum system where a single photon in a superposition of two color-states generated by a quantum emitter drives another quantum emitter with a spin-$1/2$ ground-state~\cite{Carmichael93,Gardiner93}. While heralding of the photon absorption allows us to measure classical correlations between the photonic input and the output spin states, the current success rate is not high enough to implement state transfer in a quantum network. If the success rate can be enhanced, then heralded quantum state transfer can be combined with quantum error correction. More specifically, lack of a heralding event could act as a syndrome measurement for an erasure code that could be used to increase the success rate of quantum state transfer. Moreover, extensions of this work can also apply to the realization of hybrid quantum networks connecting different physical systems via photons~\cite{Meyer15}.

The Authors acknowledge many useful discussions with M.~Kroner, W.B.~Gao and E.~Togan.
This work is supported by NCCR Quantum Photonics
(NCCR QP), the research instrument of the Swiss National
Science Foundation (SNSF), and by the Swiss NSF under
Grant No. 200020-159196.

\end{document}


\newcommand{\ddensity}[2]{\rho_{#1\,#2,#1\,#2}} 
\newcommand{\ket}[1]{\left| #1 \right>} 
\newcommand{\bra}[1]{\left< #1 \right|} 
\makeatletter \renewcommand{\thefigure}{S\@arabic\c@figure} \renewcommand{\thetable}{S\@arabic\c@table} \renewcommand{\theequation}{S\@arabic\c@equation}\makeatother

\title{SUPPLEMENTAL MATERIAL \\
Realization of a cascaded quantum system: heralded single photon absorption by a single-electron charged quantum dot}
\author{Aymeric Delteil}
\author{Zhe Sun}
\author{Stefan F\"alt}
\author{Atac Imamo\u{g}lu}
\affiliation{Institute of Quantum Electronics, ETH Zurich, CH-8093
Zurich, Switzerland.}

\maketitle

\subsection{Experimental set-up}

\begin{figure}[h]
  \centering
  \includegraphics[width=6in]{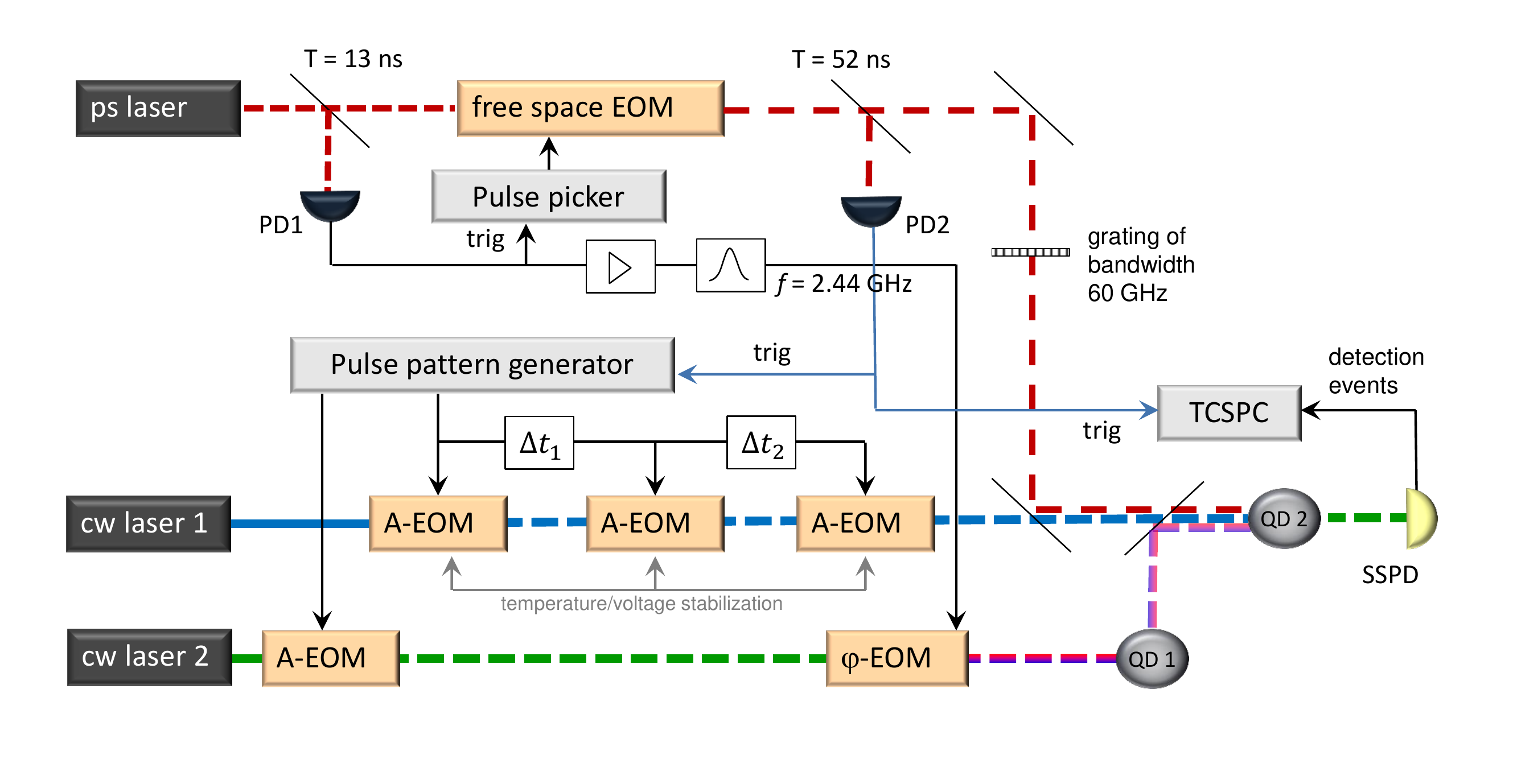}\\
  \caption{Experimental set-up. EOM: electro-optic modulator; SSPD: superconducting single-photon detector; TCSPC: time-correlated single photon counter; PD: photodiode.}\label{S1}
\end{figure}

Our experimental set-up is depicted Fig.~\ref{S1}. A mode-locked Ti-Sapphire laser produces 4-ps pulses at a rate of 76~MHz. This repetition rate is reduced to 19~MHz using a free-space electro-optic modulator (EOM) as a pulse picker. The final repetition rate is measured by a photodiode (PD2) to generate a trigger signal for both the pulse pattern generator and the time correlated single photon counter (TCSPC). The ps pulses are prolonged by a grating of bandwidth $\sim$~60~GHz and combined with the pulses produced from a CW diode laser (laser~1) using three cascaded amplitude EOMs (A-EOMs) to reach a high extinction ratio of $\sim 10^6$.
The photonic qubits are generated from QD1 using laser pulses shaped from the CW laser~2 using an amplitude EOM. Single photons at $\omega_{blue}$ and $\omega_{red}$ are generated by tuning the laser~2 at the corresponding wavelength, while single photons in superposition states are generated by tuning the laser~2 at the middle frequency $\omega_{diag}$ and modulating the output light with a phase EOM ($\varphi$-EOM) that we drive with a microwave signal at 2.44~GHz generated from filtering and amplifying the 32nd harmonic of the ps-laser repetition rate \cite{Gao12} measured by a photodiode (PD1). The amplitude of the MW signal is chosen such that the carrier amplitude vanishes. The single photon pulses emitted by QD1 are then combined with the pulses from the ps laser and from the cw laser~1. The photons scattered by QD2 are detected with a superconducting single photon detector (SSPD), and the detection events are recorded by the TCSPC triggered by the signal from PD2.

\subsection{Quantum dots with $|g_e| = |g_h|$}

\begin{figure}[h]
  \centering
  \includegraphics[width=6in]{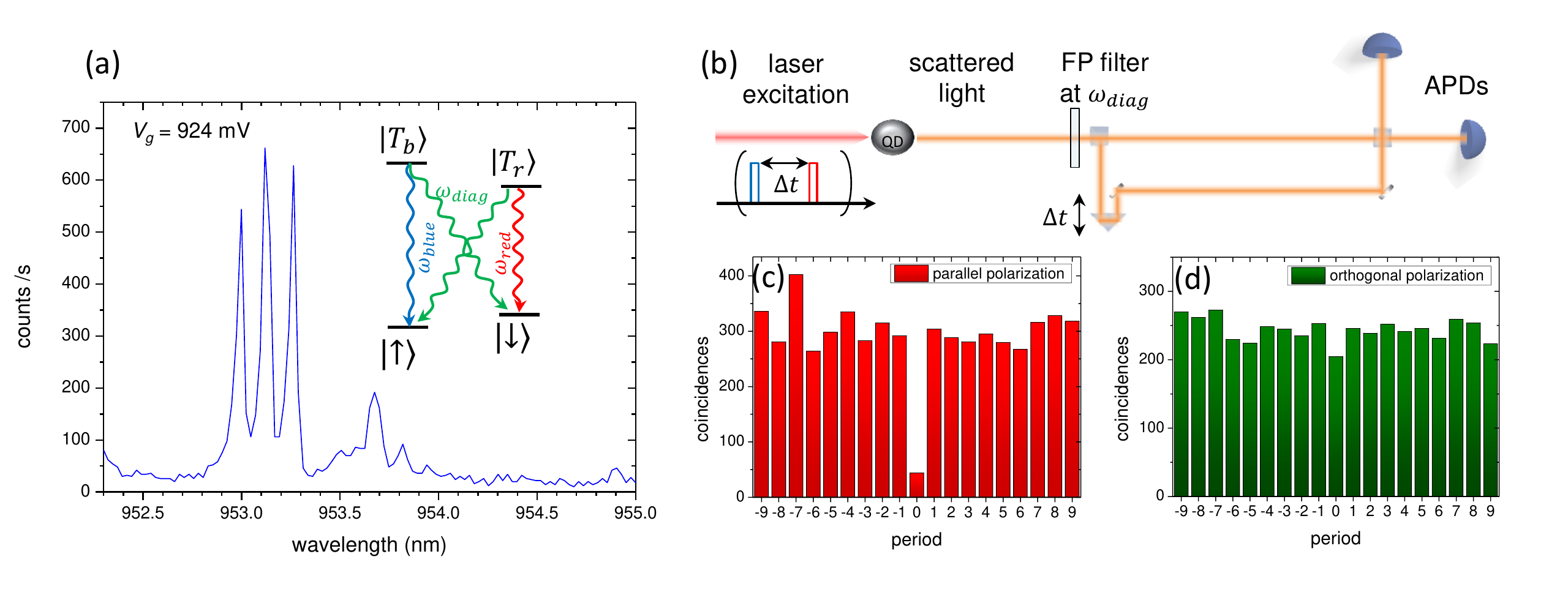}\\
  \caption{(a) Photoluminescence spectrum of the trion transitions of QD2 at 7~T. Inset: energy level scheme of QD2. (b) Set-up for the Hong-Ou-Mandel experiment, and pulse sequence consisting in alternated blue and red 400-ps laser pulses, with time difference matching the optical path length difference of the interferometer. (c) and (d) Two-fold coincidences during 4~h as a function of the period difference: (c) for parallel polarization, (d) for orthogonal polarization.}\label{S2}
\end{figure}

In self-assembled QDs the electronic in-plane $g$-factor is relatively homogeneous and usually measured around 0.5-0.6. The hole in-plane $g$-factor on the other hand originates exclusively from heavy-light hole mixing and varies strongly from one QD to another, usually spread between $0$ and $-1$. As a consequence, there is a finite probability to find QDs with close enough $|g_e|$ and $|g_h|$ such that the energy difference between the two diagonal transitions is smaller than their linewidth. To find such dots, we measured a map of our sample, with spatially resolved photoluminescence spectra, at $B = 7$~T, which is the maximum magnetic field that we can apply with our superconducting magnet. The QDs exhibiting three and only three peaks associated with the $X^-$ transitions (Fig.~\ref{S2}a shows that of QD2) are selected out and characterized using resonance fluorescence (RF), that provides a resolution better than the trion linewidth, to confirm the absence of a splitting between the two diagonal transitions (Fig.~1b of the main text shows QD2 RF plateau). Finally, a Hong-Ou-Mandel experiment is performed (Fig.~\ref{S2}b), by alternating excitation of both trions, filtering photons from the diagonal transitions and interfering them on a beam splitter. In the case of QD2, when the input polarizations are made parallel, the center peak is reduced by a factor $\sim 5$ compared to the case where the input polarizations are made orthogonal. This quantum interference of diagonal photons emitted by the two different trion states demonstrate their high degree of indistinguishability. As a consequence, no information about the initial or final electronic state can be extracted from the detection of a photon at $\omega_{diag}$. As mentioned in the main text, the $g$-factor of QDs can also be tuned using external electric, magnetic or strain field, or using optical Stark shift to obtain $|g_e| = |g_h|$

\subsection{Contributions to the background signal}

In order to faithfully attribute a detection event to a successful single photon absorption by QD2, it is necessary that the detection rate of the heralding photon be significantly higher than the background, responsible for false events. In our experiment, this background has four different origins:\\
-- dark counts from the detector\\
-- background light from the environment reaching the detector\\
-- QD light scattering caused by cw laser background (since the EOMs have finite on/off ratio, laser light passing through in the ``off" state can still scatter light)\\
-- tail of the RF decay following a QD excitation

In the first part of our work, where QD2 is driven only by light from QD1, only the two first contributions are relevant. As stated in the main text, their sum stays within 2~s$^{-1}$.

\begin{figure}[h]
  \centering
  \includegraphics[width=5in]{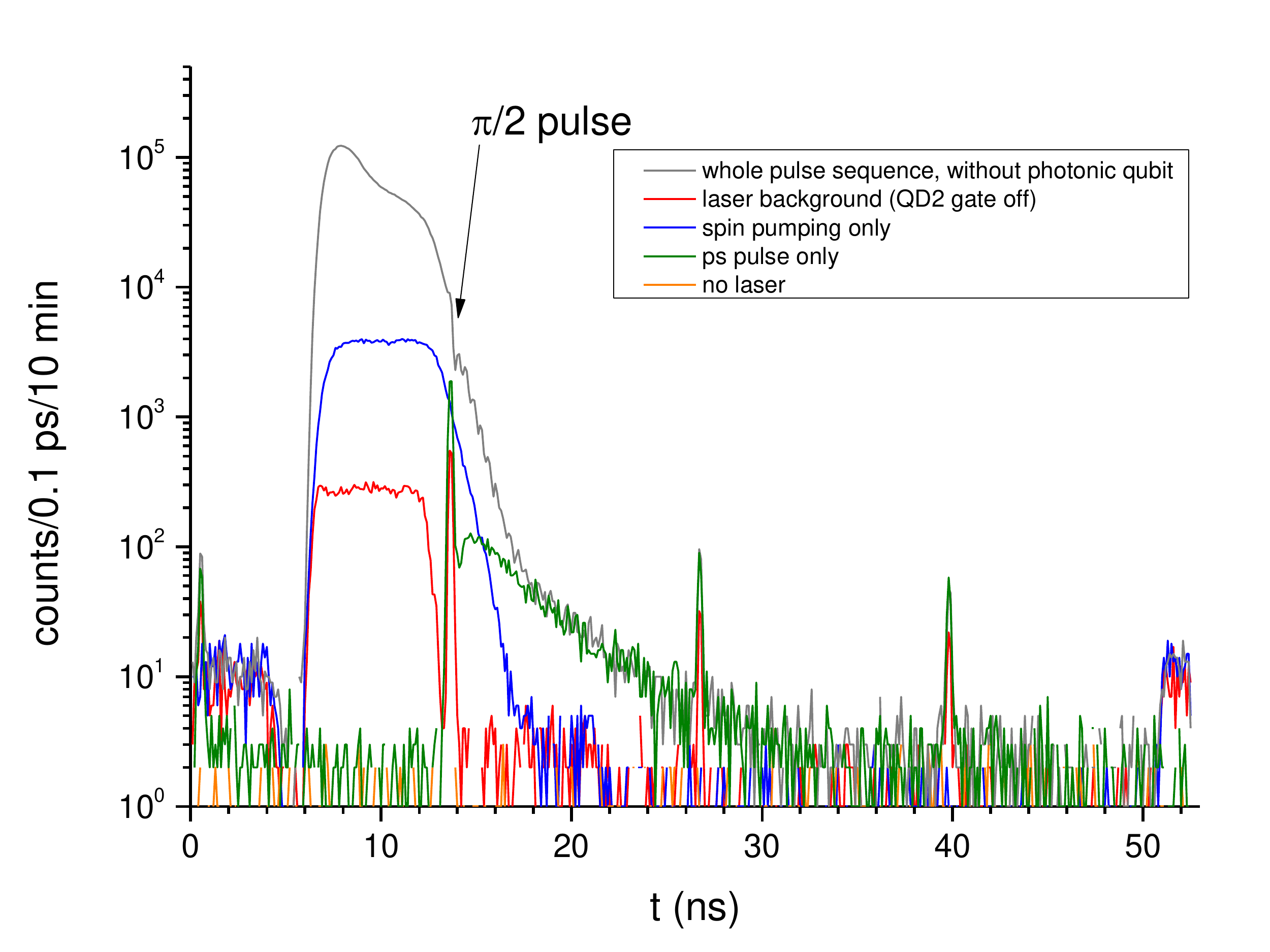}\\
  \caption{Time traces measured during 10~min, under the following conditions: gray curve: both laser on; blue curve: spin pumping laser only; green curve: ps laser only; red curve: both lasers, but QD turned off resonance; orange curve: no laser on. }\label{S3}
\end{figure}

Combining the single photon pulses with strong laser pulses that prepare, rotate and measure the spin state gives rise to the two additional contributions that henceforth dominate the background. In order to investigate them, we measured several time traces showing the different contributions, shown Fig.~\ref{S3}.
The grey curve is taken during 10~min using the laser pulse sequence described in Fig.~3a of the main text (without the photonic qubit), consisting of a spin pumping pulse followed by a rotation by $\pi/2$. It can be seen that the ratio between the maximum and the minimum count rate is about $10^5$. The decay of resonance fluorescence after the laser pulses shows a bi-exponential decay. The short decay corresponds to the excited state lifetime and is the only one that can be observed when the ps-laser is turned off (blue curve). On the other hand, the long decay can be observed when only the ps laser is left (green curve). We measured that the integrated count rate scales quadratically with the detuning as well as with the laser power, therefore we attribute this effect to relaxation following two-photon excitation in the continuum that takes place during the ps-pulse. To limit its impact we introduce a grating to prolong the ps-laser pulses. However it prevents us from introducing a spin-echo $\pi$ pulse between the $\pi/2$ pulse and the heralded absorption time window, that could cancel out the random part of the spin precession and allow us to measure quantum correlations.

The background due to imperfect polarization suppression of the laser light (red curve, measured with QD off resonance) is sizable only during the spin pumping pulse (where it is $10^2$ to $10^3$ smaller than the QD counts during this pulse as seen on the grey curve) and during the ps laser pulse, and negligibly small anywhere else. Due to imperfect suppression of the unwanted ps pulses by the free-space EOM (see Fig.~\ref{S1}), replica of the $\pi/2$ pulse at 26~ns, 39~ns and 52~ns can be seen, as well as unsuppressed echoes of the spin pumping pulse (36-44~ns, 51-52~ns and 0-4~ns), but they are all harmless as long as they do not temporally overlap with the weak signal we are interested in. Note that all these effects are usually unseen in most experiments where the background line usually lies at levels about only $10^2$ to $10^3$ smaller than the QD signal.

\subsection{Alternative protocols based on heralded single photon absorption}

The possibility to obtain QDs having $|g_e| = |g_h|$, together with our demonstration of heralded single photon absorption, opens the way for the realization of quantum information protocols based on cascaded single quantum emitters, besides the photon-to-spin transfer discussed in the main text. Here we give the outlines for realizating spin-to-spin quantum state transfer and generation of heralded entanglement between remote QD spins.

\subsubsection{Spin-to-spin quantum state transfer}

Spin-to-spin state transfer is based on photon-to-spin state transfer described in the main text, and additionally requires spin-to-photon conversion as a first step. The process is depicted Fig.~\ref{S4}a: the first QD is in the single charge state, in Voigt geometry. Its transitions from $|\uparrow \rangle $ to the trion states match the bluemost and redmost transitions of the destination QD.  Starting from a spin state $\alpha |\uparrow \rangle + \beta |\downarrow \rangle$, we excite the transions $|\uparrow \rangle \rightarrow |T_b \rangle$ and $|\downarrow \rangle \rightarrow |T_r \rangle$ with optical $\pi$-pulses. Amongst the possible radiative decay channels, we select using filters the ones leading to the same final spin state $|\uparrow \rangle $, to prevent entanglement of the emitted photon with the spin degree of freedom (post-selection of heralded events will ensure that the final spin state is $|\uparrow \rangle $), and we erase the polarization degree of freedom with a polarizer. The photon transmitted in the channel is then in the state $\alpha |\omega_{blue} \rangle + \beta |\omega_{red}  \rangle$ and can be then transferred to the spin of the second dot.

\subsubsection{Generation of heralded spin entanglement}

\begin{figure}[h]
  \centering
  \includegraphics[width=5in]{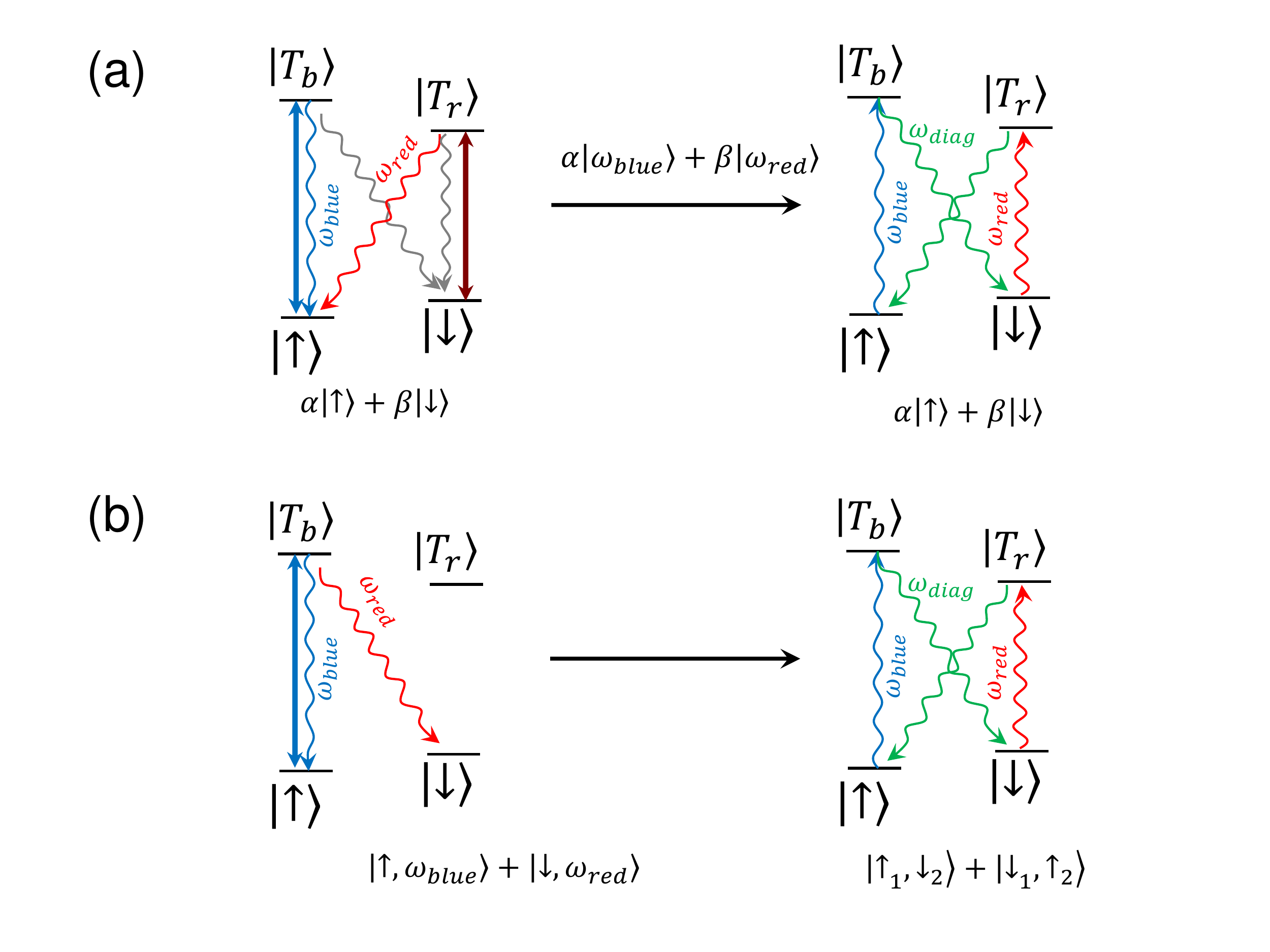}\\
  \caption{(a) Spin-to-spin state transfer. The decay channels of the first dot shown in grey are filtered out. (b) Generation of heralded spin entanglement using heralded single photon absorption}\label{S4}
\end{figure}

Generation of distant spin entanglement can be implemented by replacing the photonic qubit sent to the destination QD by a photon whose color degree of freedom is entangled with the spin of the initial QD (Fig.~\ref{S4}b) Generation of such spin-photon entangled pair can be done as described in~\cite{Gao12}, by excitation of a single trion state. In Voigt geometry where the trion state decays to the two spin ground states with equal probability, the emitted photon is maximally entangled with the spin state. After having erased the polarization degree of freedom, the spin-photon entangled state takes the form $|\uparrow, \omega_{blue} \rangle + |\downarrow, \omega
_{red}\rangle$. Upon heralded absorption of the photonic part, the two-QD spin state is left in a maximally entangled state of the form $|\uparrow_1, \downarrow_2 \rangle + |\downarrow_1, \uparrow_2 \rangle$

{}

\clearpage